\documentclass[12pt]{iopart}
\usepackage{graphicx}

\newcommand{\ssT}{{\scriptscriptstyle T}}
\newcommand{\pT}{p_\ssT}
\newcommand{\Comment}[1]{}
\usepackage{iopams}  
\usepackage{color}

\begin{document}

\title[Search for a Ridge Structure Origin with Shower Broadening and Jet Quenching]{Search for a Ridge Structure Origin with Shower Broadening and Jet Quenching}

\author{R. Mizukawa$^1$, T. Hirano$^2$, M. Isse$^3$, Y. Nara$^4$, A. Ohnishi$^{1,5}$
}

\address{
1 Department of Physics,
Faculty of Science, Hokkaido University,\\~~ Sapporo 060-0810, Japan}
\address{2
Department of Physics, Graduate School of Science, University of Tokyo,\\
~~Tokyo 113-0033, Japan}
\address{3 Kobe City College of Technology, Kobe 651-2194, Japan}
\address{4 Akita International University, Akita 010-1211, Japan}
\address{5 Yukawa Institute for Theoretical Physics, Kyoto University,
Kyoto 606-8502, Japan}
\ead{rei@nucl.sci.hokudai.ac.jp}

\begin{abstract}
We investigate the role of jet and shower parton broadening
by the strong colour field
in the $\Delta\eta$-$\Delta\phi$ correlation of high $\pT$ particles.
When anisotropic momentum broadening ($\Delta p_z > \Delta\pT$) is given to
jet and shower partons in the initial stage,
a ridge-like structure is found to appear in the two hadron correlation.
The ratio of the peak to the pedestal yield is overestimated.
%
\end{abstract}


\section{Introduction}

RHIC experiments have offered us rich information on hot and dense matter.
One of the most surprising phenomena may be the ridge structure
observed in the $\Delta\eta$-$\Delta\phi$ correlation~\cite{ridge}.
Two high $\pT$ particles are found to have a base-like (pedestal) correlation
elongated in the $\Delta\eta$ direction
in addition to the normal jet peak correlation
having narrow $\Delta\eta$ and $\Delta\phi$ width.
We cannot explain the ridge structure in a standard hadronisation mechanism,
where high $\pT$ hadrons are formed from jet fragmentation
and correlated hadrons have similar rapidities to jet.
Thus the ridge structure indicates the correlation
between a jet and other particles, which form high $\pT$ hadrons.
Furthermore,
the correlation should be generated in the early stage,
where two particles having a large rapidity gap are still spatially close.

There are several proposals on the origin of
the ridge structure~\cite{Shuryak,Wong,Majumder,Schenke}.
Here we concentrate on the momentum broadening
from glasma~\cite{Majumder,Schenke}.
When two colour glass condensates (CGC) collide
and the system created by such a collision goes to a thermal state (QGP),
we expect the existence of strong colour field before thermalisation
which we call glasma. As the system expands, colour field will decay
and becomes weak at the time scale of $\tau\sim 1/Q_s$,
where $Q_s$ is the saturation scale.
Due to instabilities of classical colour fields~\cite{Schenke,Dumitru}, 
colour fields, however, may grow exponentially
and would cause momentum broadening of jet and shower partons.

In this work, we investigate the effects of the momentum broadening
in the early times on the {\em hadronic} $\Delta\eta$-$\Delta\phi$ correlation.
For this purpose, we also consider later processes
in the framework of the jet-fluid string (JFS) model~\cite{Isse07},
which treats energy loss of jet partons in the QGP phase,
hadronisation through formation of a string from a jet parton and a parton 
in a fluid element at the surface of the QGP phase and decay of it.

\section{Jet-Fluid String model with Initial Momentum Broadening}

For the description of high $\pT$ hadron production,
we start from the JFS model~\cite{Isse07},
which explains various high $\pT$ QGP signals,
such as jet quenching, large elliptic flow at high $p_T$, 
and disappearance of the backward azimuthal angle correlation simultaneously.
Momentum broadening is assumed to act on partons just after the jet production.
In total, 
we take the following four stages into account.
(1) {\bf Jet production}: 
In the initial stage, 
hard processes create jet partons and several shower partons are emitted
from them~\cite{PYTHIA}.
(2) {\bf Momentum broadening}:
Momentum distribution for these partons is largely broadened
in the initial strong colour fields, namely glasma.
(3) {\bf Energy loss in QGP}:
When a system reaches a thermal state,
jet and shower partons propagate in QGP and lose their energy~\cite{GLV}.
We describe space-time evolution of QGP fluids using full three dimensional
ideal hydrodynamics~\cite{Hirano}.
(4) {\bf Hadronisation}:
In the final state, we assume that jet and shower partons
form strings with fluid partons.
High $\pT$ hadrons are formed from the decay of these strings.

Broadening in glasma has the following characteristic features.
Firstly, the momentum transfer in hard processes
is much larger than the saturation scale,
then the transverse interaction area is smaller
than the typical domain size $1/Q_s^2$.
Secondly, a colour force gives momentum kicks in opposite directions
to partons having opposite colour charges.
Thirdly, the field strength would be stronger in the longitudinal directions
than in the transverse direction
up to the time scale of $\tau\sim 1/Q_s$~\cite{Lappi,Schenke}.
Therefore, we expect larger broadening in the $p_z$ direction.
The same conclusion was drawn
under the assumption of random transversely polarized colour-magnetic
fields~\cite{Majumder}.
Thus we model the broadening in glasma,
as the opposite random Gaussian kick for the two end points
in the case of a simple $q\bar{q}$ initial string,
and the longitudinal kick width ($\Delta p_z$) is assumed to be larger
than the transverse kick width ($\Delta \pT$).

\begin{figure}[t]
\begin{center}
\includegraphics[width=8.0cm]{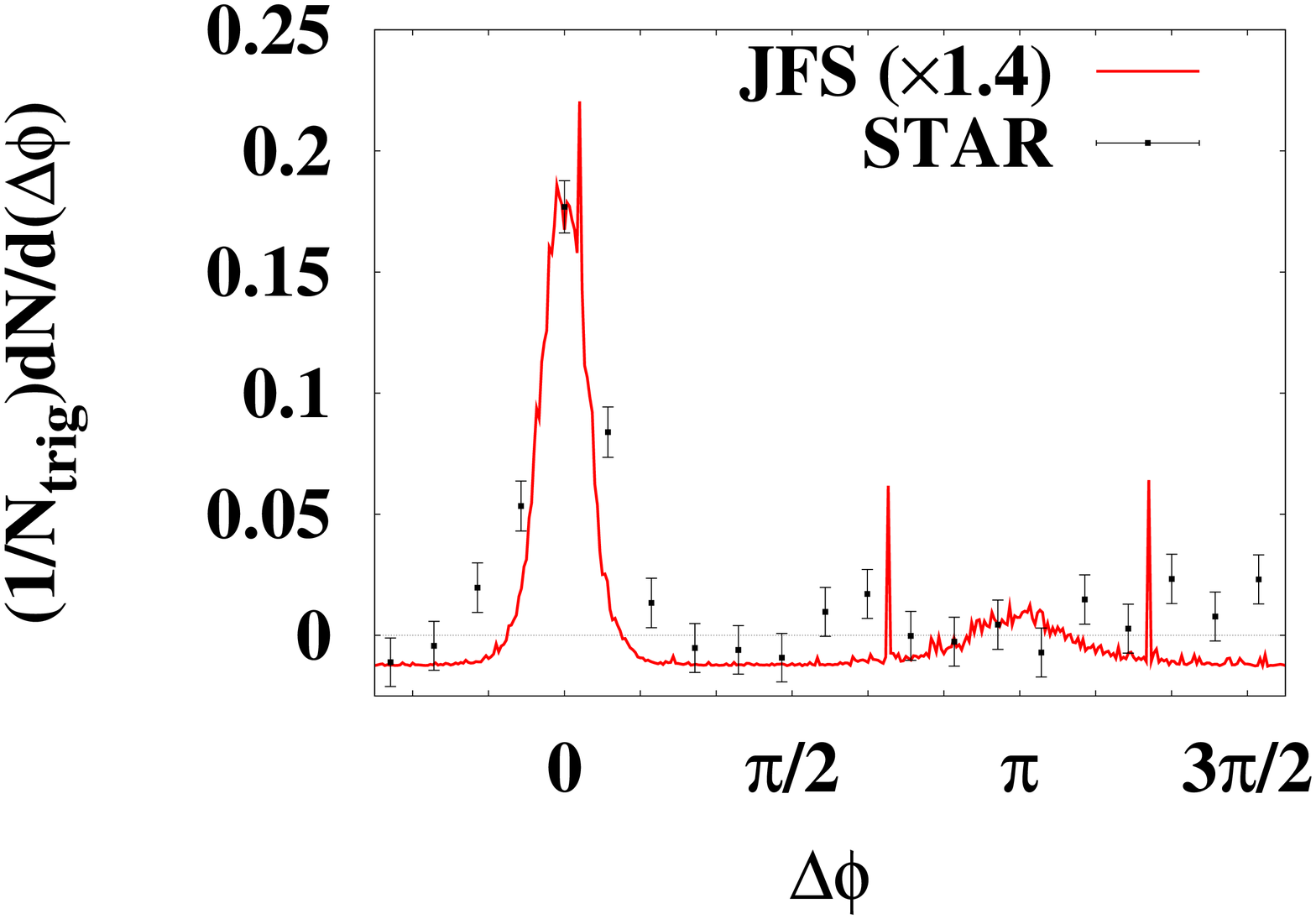}~%
\includegraphics[width=8.0cm]{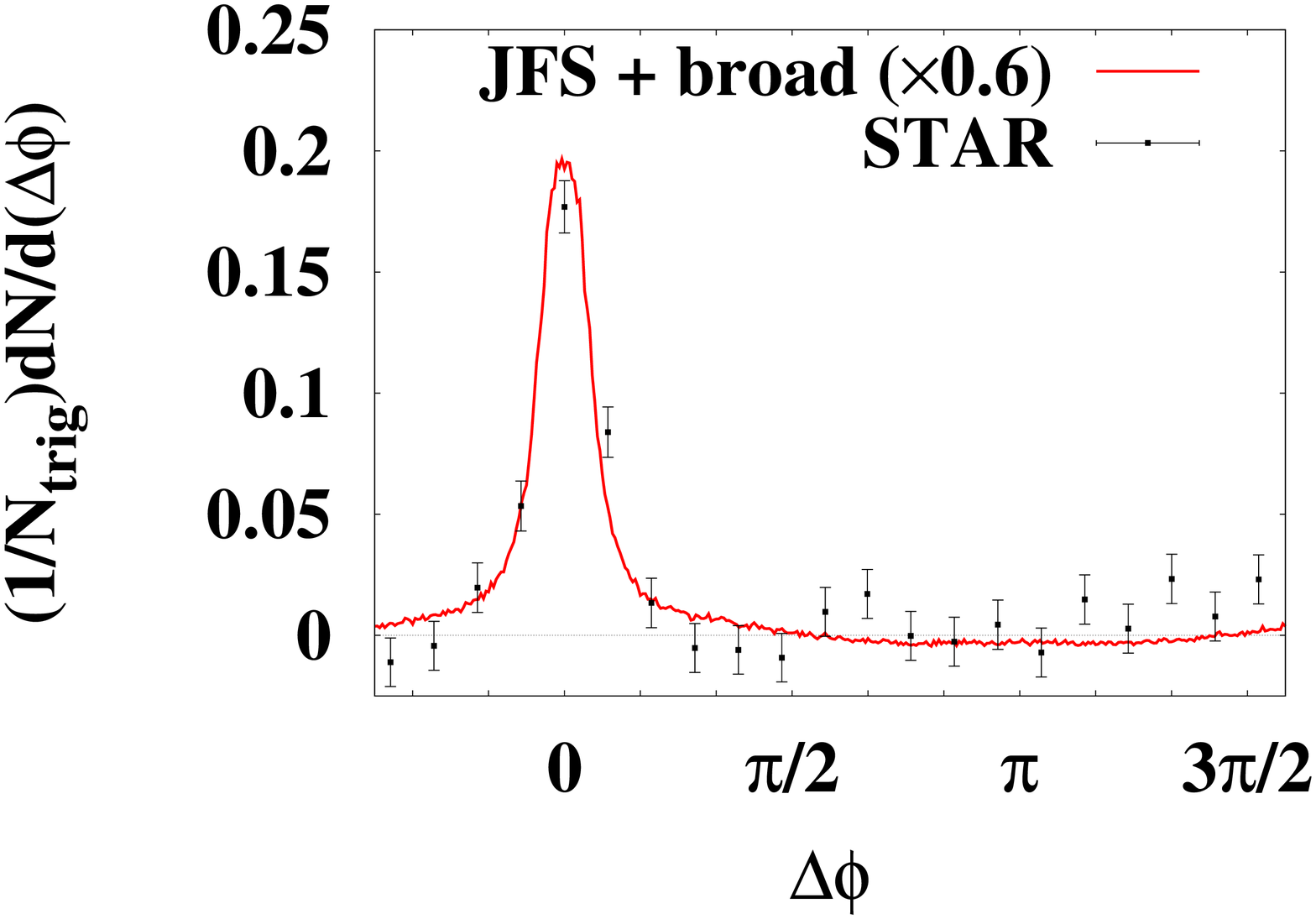}
\caption{Azimuthal angle correlation in JFS without (left panel)
and with (right panel) momentum broadening.}
\label{Fig:phi}
\end{center}
\end{figure}

\section{Results} 

In Fig.~\ref{Fig:phi}, we show the two hadron azimuthal angle correlation
calculated without (left) and with (right) broadening.
We have adopted the momentum broadening widths
$(\Delta p_z, \Delta\pT)=(10~\mathrm{GeV}/c, 2.5~\mathrm{GeV}/c)$
as a typical parameter set.
We find that the backward peak disappears in both cases.
Without broadening,
the absolute yield of the near side correlation is underestimated.
This may be a natural consequence of the missing pedestal part
in the $\Delta\eta$-$\Delta\phi$ correlation.
With momentum broadening, the near side peak has a little larger width
and a larger yield compared with those without broadening.
The enhanced yield mainly comes from shower partons.
Shower partons generally have small $\pT$ and most of them cannot make
high $\pT$ hadrons without momentum broadening, 
but they can survive in traversing QGP and make high $\pT$ hadrons
when the effect of transverse momentum broadening is enough.

\begin{figure}[tb]
\begin{center}
\includegraphics[width=8.5cm]{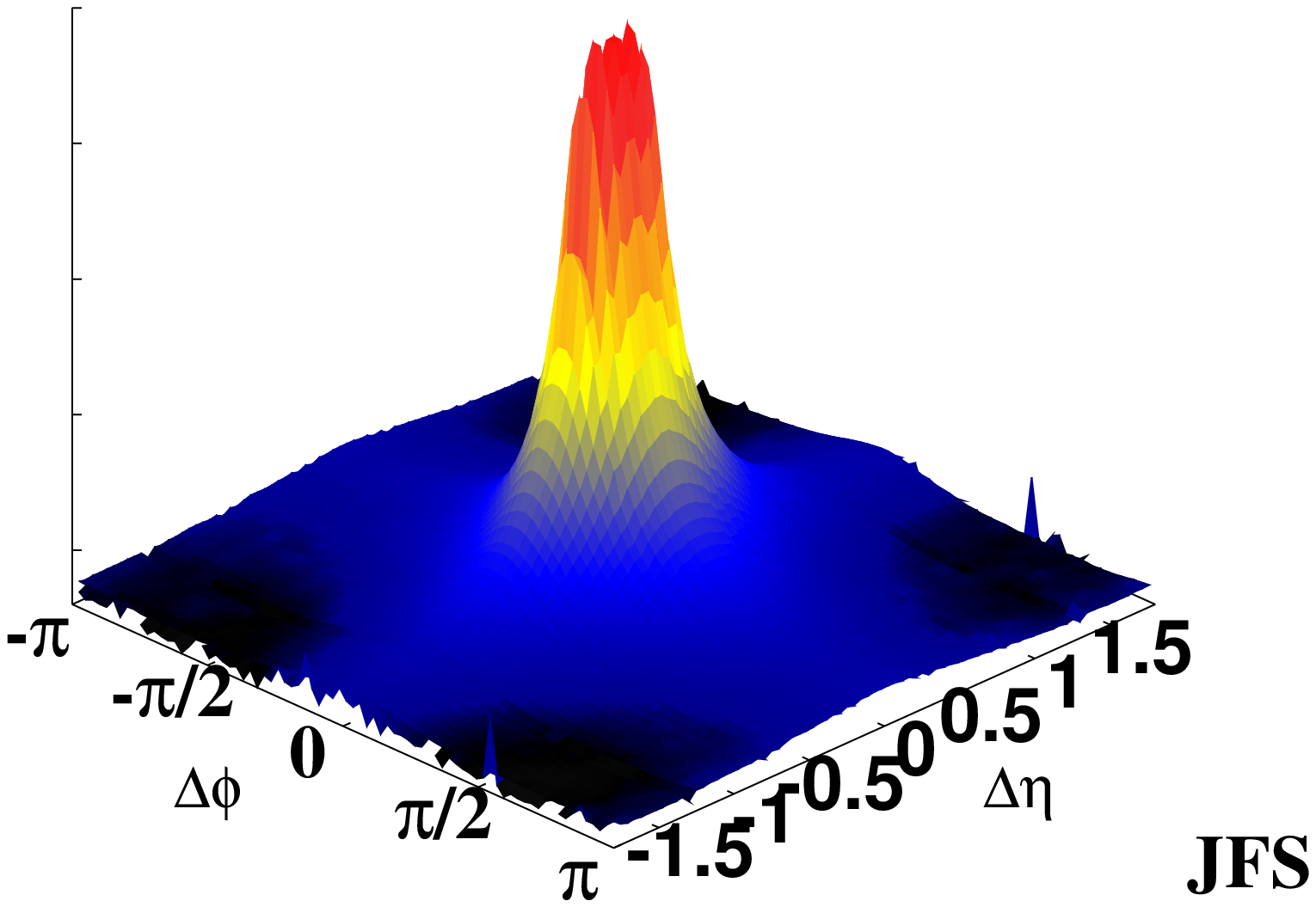}~%
\includegraphics[width=8.5cm]{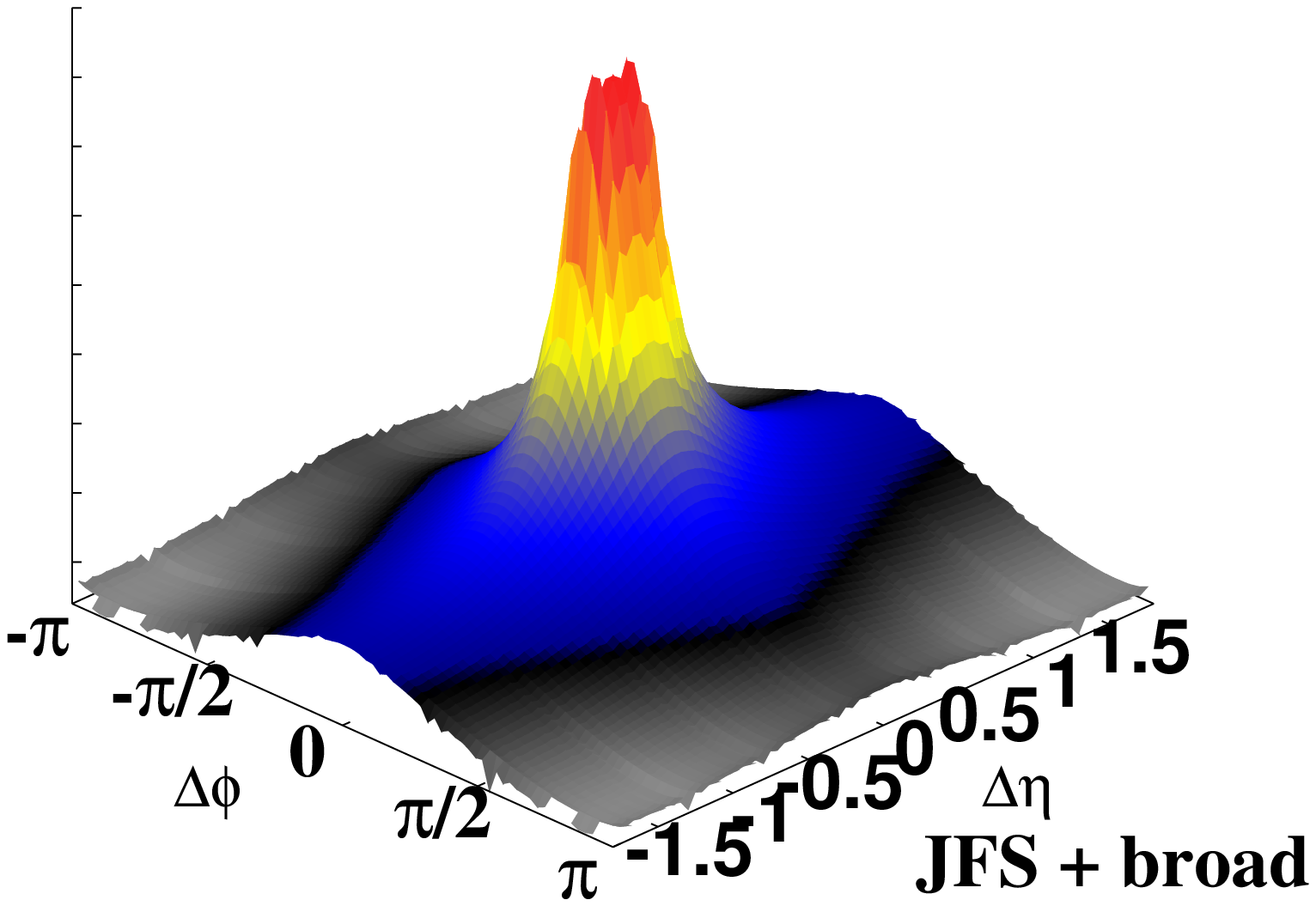}
\caption{
Calculated $\Delta\phi$-$\Delta\eta$ correlation without (left panel)
and with (right panel) momentum broadening.
}
\label{Fig:ridge}
\end{center}
\end{figure}

In Fig.~\ref{Fig:ridge}, 
we show the results of $\Delta\eta$-$\Delta\phi$ correlation
without (left) and with (right) momentum broadening.
JFS without momentum broadening does not show the ridge structure.
While a jet-fluid string has two end points having different rapidities,
the rapidity gap between the jet and fluid partons is not large
and fluid partons have small $\pT$.
As a result, all high $\pT$ ($\pT > 2~\mathrm{GeV}/c$) hadrons
from JFS decay have similar rapidities to the jet parton.
Whereas, a ridge-like structure appears when we implement momentum broadening.
In this result, 
the peak and pedestal parts have different origins;
hadrons in the peak are formed from the jet-fluid string fragmentation,
but the pedestal part is constructed by the shower-fluid string fragmentation.
From the calculated results with different parameter sets,
we find that $\Delta p_z \gtrsim 5~\mathrm{GeV}/c$
and $2.5~\mathrm{GeV}/c \lesssim \Delta\pT < \Delta p_z$ is the condition
for the ridge-like structure to appear,
and is consistent with the estimate
with instability~\cite{Schenke}.
It would be natural that large longitudinal broadening is preferred
to make an elongated tail in the $\Delta\eta$ direction.
It should be noted that transverse broadening is also necessary
for shower partons to make high $p_T$ hadrons,
as discussed in $\Delta\phi$ correlation.
But this $\Delta p_T$ should not be too much
to keep small $\Delta\phi$ width.

\section{Summary}

In this work, we have discussed the effects of
momentum broadening before thermalisation
as the origin of the ridge structure.
This kind of broadening may be generated by the glasma.
We have given the momentum broadening to the jet and shower partons
in the pre-equilibrium stage,
and evaluated the hadron correlation
after energy loss in QGP and hadronisation processes
through string fragmentation.
We have found that, if we have anisotropic momentum broadening,
strings from jet and shower partons may make the ridge-like structure.
Therefore, momentum broadening in glasma is a possible origin
to create the ridge structure.

In this work, the momentum broadening strengths are taken as free parameters,
which is independent from the position and momentum.
This treatment corresponds to a constant colour electric field for one jet.
In addition, the ratio of the peak to the base part in the ridge is too large.
Further investigations in these directions are necessary
for a deeper understanding of the ridge.

This work is supported in part
by the Ministry of Education,
Science, Sports and Culture, Grant-in-Aid for Scientific Research
under the grant numbers,
    1707005 (AO),		
    19540252 (AO),		
and 
    19740130 (TH).		

\end{document}